\documentclass{elsart}
\usepackage{graphicx,amssymb}
\journal{Astroparticle Physics}

\begin{document}
\begin{frontmatter}
\title{Detection possibility of the pair-annihilation neutrinos from the neutrino-cooled pre-supernova star}
\author[ifuj]{A. Odrzywolek},
\ead{odrzywolek@th.if.uj.edu.pl}
\author[ifuj]{M. Misiaszek\corauthref{cor}},
\ead{misiaszek@zefir.if.uj.edu.pl}
\corauth[cor]{Corresponding author.}
\author[ifuj,ifj]{M. Kutschera}
\ead{Marek.Kutschera@ifj.edu.pl}
\address[ifuj]{M. Smoluchowski Institute of Physics, Jagiellonian University, Reymonta 4, 30-059 Krakow, Poland}
\address[ifj]{H. Niewodniczanski Institute of Nuclear Physics, Radzikowskiego 152, 31-342 Krakow, Poland}

\begin{abstract}
The signal produced in neutrino observatories by the pair-annihilation neutrinos emitted from
a 20~$M_{\odot}$ pre-supernova star
at the silicon burning phase is estimated. The spectrum of the neutrinos with an average energy
$\sim$2~MeV is calculated with the use of
the Monte Carlo method. A few relevant reactions for neutrinos and anti-neutrinos in modern detectors are considered.
The most promising results are from $\bar{\nu}_e + p \longrightarrow n + e^{+}$ reaction.
During the Si-burning phase we expect 1.27 neutrons/day/kiloton of water to be produced by
neutrinos from a star located at a distance of 1~kpc.
Small admixture of effective neutron-absorbers as e.g. NaCl or GdCl$_{3}$ makes these neutrons
easily visible because of Cherenkov light produced by electrons which were hit by
$\sim$8~MeV photon cascade emitted
by Cl or Gd nuclei.
The estimated rate of neutron production for SNO and Super-Kamiokande is, respectively, 2.2 and 41 events
per day for a star at 1~kpc. For future detectors UNO and Hyper-Kamiokande  we expect 5.6 and 6.9 events
per day
even for a star 10~kpc away. This would make it possible to foresee a massive star death
a few days before its core collapse. Importance of such a detection for theoretical astrophysics is
discussed.
\end{abstract}
\end{frontmatter}

\section{Introduction}
The detection of solar neutrinos by the R.~Davis's chlorine detector and
supernova 1987A neutrinos by the Kamiokande water Cherenkov observatory are
milestones of modern science, {\it Nobel Prize} honored in 2002. Currently built and operating
neutrino detectors are larger, much more efficient and are able to detect all neutrino species in a wide
range of energy. There are already proposed new experiments with
next-generation megaton-scale detectors, like Hyper-Kamiokande \cite{hk} and UNO \cite{uno}.

\begin{table}[b]
\begin{tabular}{ c c c }
\hline
& The Sun & Neutrino-cooled phase of 20~M$_\odot$ star \\
\hline
Lifetime & $10^{10} $ years & 300 years \\
Photon luminosity & L$_\odot$ & 10$^5$L$_\odot$ \\
Maximum $\nu$ luminosity & 0.02 L$_\odot$ & 10$^{12}$ L$_\odot$ \\
Average neutrino energy & 0.3 MeV & 0.7-2 MeV \\
Total $\nu$ energy released & 10$^{49}$ erg & 10$^{51}$ erg \\
\hline
\end{tabular}
\caption{Differences between the Sun (main sequence star) and the neutrino-cooled
20~M$_\odot$ star (\cite{woosley}, Table~1). Approximate values (orders of magnitude) are given.
The total energy
carried by neutrinos for a 20~M$_\odot$ neutrino-cooled star  corresponds only to late stages of
nuclear burning (without supernova neutrinos).}
\label{table1}
\end{table}

Solar neutrinos were detected with sub-kiloton detectors, because of relatively small distance to
the Sun and continuous
emission. Supernova neutrinos carry enormous energy of 10$^{53}$ ergs released in gravitational collapse
of the stellar core.
This makes their detection possible from quite large distances, but such events are rare.
We consider here the feasibility  of detection of
another giant astrophysical sources of neutrinos, namely massive (M$>$8M$_\odot$) stars at late stages of
their evolution. Such stars, during C, Ne, O and Si burning phases, emit neutrinos copiously and are
referred to as neutrino-cooled stars (\cite{arnett}, Sect. 10)  or
pre-supernova stars \cite{langanke}. The structure of such evolutionary advanced stars is
different from main sequence stars like
the Sun. The most important differences relevant to the neutrino detection possibility are
summarized in Table~\ref{table1}.

The solar neutrino luminosity is $7.8 \cdot 10^{31}$ erg/s \cite{bahcall}.
The maximum neutrino luminosity during Si burning exceeds
$3 \cdot 10^{45}$ erg/s  (\cite{woosley},~\cite{zimmerman})
-- a value larger by a factor of $3.85 \cdot 10^{13}$. The neutrino
energy flux at Earth from such a star
and from the Sun are equal for a star
$\sqrt{3.85 \cdot 10^{13}} = 6.2 \cdot 10^6 $AU, i.e. 30~pc away. This indicates that neutrino-cooled stars could
be detected at astronomical distances. Actually, because of
a different spectrum (cf.~Fig.~\ref{solar})
and the presence of anti-neutrinos (cf. Table~\ref{f_factor}),
we are able to detect them from kiloparsec distances.

Massive (neutrino-cooled) stars are believed to end as
{\it core-collapse} supernovae. SN~1987A confirmed
this theory. This close connection may be used to estimate
chance of neutrino-cooled stage observation.
If some of massive stars die other way, rate of
neutrino-cooled events may be higher than of SNe, but
it is reasonable to use estimates made for supernovae. At least
three different methods exist there:
historical records \& remnants counts \cite{vandenbergh},
extragalactic observations \cite{capellaro_turrato} and
population synthesis (i.e. simulated
stellar evolution of the Galaxy) methods  \cite{piran}. Only first
method gives {\it local rate} directly: time between
events less than 5kpc away is estimated to be 175 years \cite{strom}.
Two other methods give rate for entire Galaxy.
Extragalactic counts give 40\ldots 200 years \cite{capellaro_turrato}
of average time interval between supernova explosions.
Population Synthesis give 10 years \cite{piran}.
To compare {\it local rates} we may use Galaxy model of e.g
Bahcall\&Soneira \cite{galaxy_model}. For the solar neighbourhood
we found 0.5\%, 10\% and 50\% of disc stars closer than 1kpc,
5kpc and 10kpc respectively.
Assuming that the Sun is not in privileged position close to
e.g. giant star forming region, we get
time between supernovae closer than 5kpc from 100 (Population
Synthesis) to 400\ldots2000 (extragalactic counts) years.
Apparent disagreement between these two methods is usually explained
by a large number of unobserved events. Clouds of interstellar
gas and dust obscuring optical detection are obviously transparent
for neutrinos, so we may conclude that neutrino-cooled
events closer than 5kpc may be observed even more frequent than one
per century.
Close candidates for such a detection apparently do exist,
with Betelgeuse ($\beta$~{\it Ori}) being the most popular example
\cite{apod}. In Sect.~5 we
combine rates discussed above with our results to estimate a chance
of successful detection.

As an example we consider a 20~M$_\odot$ star model with properties reported in Table~1 of Ref. \cite{woosley},
with {\it explicite} given total neutrino luminosities of various stages of nuclear burning. As our scope here
is to study the feasibility of neutrino detection, not to find state-of-art details of emission\footnote{
Detailed picture of late stages of nuclear burning is still a subject
of active research as there are unanswered questions about e.g. hydrodynamical nature of nuclear burning
\cite{2dburning}. Neutrinos itself are still enigmatic objects and e.g. the role of
spin-flip interactions and the influence of right-handed neutrinos~\cite{rightnu} is investigated.
},
we assume that the entire neutrino flux is resulting from $\mathrm{e}^+\mathrm{e}^-$ annihilation.
The values from Table~\ref{table2} of the temperature, density and the electron chemical potential at the stellar
center are assumed to approximate conditions in the stellar core.
For simplicity we treat neutrino propagation without oscillation, however 
the effect of neutrino oscillations may change significantly the number of observed
events in a given experiment \cite{rubbia}.
To calculate the relevant neutrino cross sections the standard Weinberg-Salam theory \cite{gws} is used.

In Sect.~2 we calculate the neutrino spectrum with the use of Monte Carlo method, in Sect.~3 we briefly present
neutrino-induced reactions and outline of selected detectors.  Sect.~4 contains the expected
neutrino signal from a star located at a distance of 1~kpc. In Sect.~5  astrophysical
implications of the pre-supernova neutrino detection are discussed.

\begin{table*}
\begin{tabular}{ c c c c c c c }
\hline
Burning  & $T_c$ & $\rho_c$ & $\mu_e $ & $L_{\nu}$ & Duration & Total energy \\
 Phase &[MeV]& [g/cc] & [MeV] & [erg/s] &  $\tau$ & emitted [erg] \\
\hline
C  & 0.07 &$2.7 \cdot 10^5$& 0.0& $7.4\cdot 10^{39}$ & 300 yrs & $ 7 \cdot 10^{49} $ \\
Ne & 0.146 &$4.0 \cdot 10^6$& 0.20& $1.2 \cdot 10^{43}$ & 140 days & $1.4 \cdot 10^{50}$\\
O   & 0.181 &$6.0 \cdot 10^6$& 0.24& $7.4 \cdot 10^{43}$ & 180 days& $1.2 \cdot 10^{51} $\\
Si &   0.319  &$4.9 \cdot 10^7$&  0.84 & $3.1 \cdot 10^{45}$ & 2 days & $5.4 \cdot 10^{50}$\\
\hline
\end{tabular}
\caption{Properties of a 20~M$_\odot$ star according to Ref.~\cite{woosley}. We have calculated the total
energy radiated in neutrinos as a product $\tau L_{\nu}$. Actually, the neutrino emission is expected
to be a function of time.}
\label{table2}
\end{table*}

\section{The spectrum of pair-annihilation neutrinos}

Neutrinos produced by thermal processes are the most important part of the neutrino flux balancing
the nuclear energy generation in the central region of massive (M$>$8M$_{\odot}$) stars
at late phases of nuclear burning, i.e. from carbon burning  up to silicon burning.
For simplicity, we have assumed that the entire neutrino flux is produced by pair
annihilation.\footnote{
Inspection of Fig.~3, 4, 8, 9, 12, and 13 from Itoh et. al. \cite{itoh} together with the values of
electron fraction, $Y_e=0.5$, central
temperature, $T_c$, and density, $\rho_c$,
from Table~\ref{table2}
shows that the annihilation neutrinos are dominant.}
Actually, so-called photoneutrinos
and neutrinos from plasmon decay (cf. \cite{aufderheide}, Fig.~1) may contribute to the total neutrino
flux \cite{itoh}, depending on physical conditions.
This assumption  is valid up to the silicon burning. After this phase the amount
of neutrinos produced by weak nuclear reactions (beta decays, electron capture)
increases, and finally dominates the neutrino flux \cite{langanke}.

Pair annihilation neutrinos are produced in the reaction (\cite{kippenhahn}, p.~171)
\begin{equation}
\label{eplus_eminus}
e^{+} + e^{-} \longrightarrow {\nu}_x +{ \bar{\nu}}_x
\end{equation}

A high temperature (T$>{10}^9$K)  is required to produce enough $e^{+} e^{-}$ pairs. Usually, in the local
thermodynamical equilibrium,
these pairs annihilate back into photons, but sometimes the reaction (\ref{eplus_eminus}) occurs.
Neutrinos produced by the reaction (\ref{eplus_eminus}) escape freely from the central region of a star. In reaction
eq.~(\ref{eplus_eminus}) the fluxes of neutrinos and anti-neutrino are the same.

We calculate the spectrum of neutrinos produced in reaction (\ref{eplus_eminus})
with the Monte Carlo method of Ref. \cite{fuller}.
Both, electrons and positrons are described by Fermi-Dirac (FD) distributions.
Conditions in the central region of the 20~M$_\odot$ star which define FD distribution parameters
are summarized in Table~\ref{table2}.

In the simulation we pick up electron and positron four-momenta from FD distributions,
transform to the center-of-mass frame,
distribute neutrino momentum directions randomly, and convert neutrinos energy back
to the rest frame. Every single event is binned and counted as $|M|^2$, where our annihilation matrix is
proportional to:
\begin{eqnarray}
{|M|}^2 \propto&
 (C_A - C_V)^2 (   p_{e^{-}} \cdot q_{{\nu}_{x}}) (p_{e^{+}}  \cdot q_{{\bar{\nu}}_{x}})&+\\ \nonumber
 &(C_A + C_V)^2 (   p_{e^{+}} \cdot q_{{\nu}_{x}}) (p_{e^{-}} \cdot q_{{\bar{\nu}}_{x}})&+\\
&{m_e}^2 ({C_V}^2-{C_A}^2) q_{{\nu}_{x}} \cdot q_{{\bar{\nu}}_{x}} & \nonumber
\end{eqnarray}
Here $p$ and $q$ are four-momenta, $m_e$ is the electron mass and:
\begin{equation}
C_V = \frac{1}{2} \pm 2\; {\sin}^2 {\theta}_W \, , \qquad
C_A = \frac{1}{2},
\end{equation}
where $ {\theta}_W$ is the Weinberg angle and $ {\sin}^2 {\theta}_W = 0.2224$. The `$+$' sign  refers to
electron neutrinos,
while the `$-$' sign is for $\mu$ and $\tau$ neutrinos. The relative
number of events from two simulation runs (one for the `$+$' sign, the other one for the `$-$'sign) has
been used to determine the $\nu_{\mu,\tau}/\nu_e$ ratio.

\begin{figure}
\center
\includegraphics[angle=0,scale=.5]{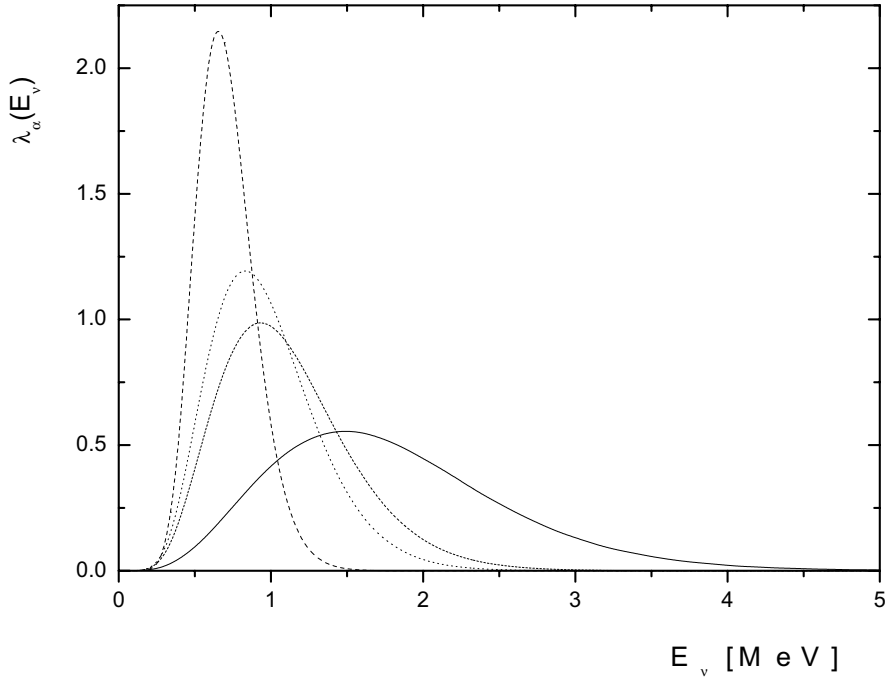}
\caption{Normalized spectrum of pair-annihilation anti-neutrinos emitted by 20~M$_\odot$ star during
carbon (dashed line), neon (dotted line), oxygen (short-dashed line) and silicon (solid line) burning stage.
The spectrum shape for all flavors
of neutrinos and anti-neutrinos is very similar.
However, relative fluxes of given neutrino flavors (cf.~Table~\ref{f_factor}) are different.
There is equal amount of neutrinos and anti-neutrinos.
The average anti-neutrino energy is 1.80 MeV, 1.11 MeV,
0.97 MeV, 0.71 MeV for Si, O, Ne, and C
burning stage, respectively.
As we expect, for $T\rightarrow 0$ the spectrum shape approaches the  $E_{\bar{\nu}}=m_e=$0.51 MeV annihilation line.}
\label{spectra}
\end{figure}

The resulting anti-neutrino spectrum is presented in Fig.~\ref{spectra}.
We have used this spectrum together with the neutrino luminosity from Table~\ref{table2} and
the flavor fractions given in Table~\ref{f_factor} to find
the detector response.

\begin{table}
\begin{tabular}{ c c c c }
\hline
Burning phase & $\nu_e$ ($\bar{\nu}_e$) fraction & $\nu_{\mu,\tau}/\nu_e$ ratio & Average $\nu_x$ energy \\
\hline
C 	& 42.5 \%  & 1:11.4  & 0.71 MeV \\
Ne	& 39.8 \%  & 1:7.8    & 0.99 MeV \\
O	& 38.9 \%  & 1:6.9    & 1.13 MeV \\
Si	& 36.3 \%  & 1:5.4    & 1.85 MeV \\
\hline
\end{tabular}
\caption{Fraction of given neutrino flavor emitted by pair-annihilation, used in formula (\ref{r}). One can notice
increasing with temperature fraction of muon and tau neutrinos.}
\label{f_factor}
\end{table}

\section{Neutrino reactions and cross-sections in modern neutrino detectors}

\subsection{Brief description of selected experiments}

Super-Kamiokande \cite{sk} and SNO \cite{sno} are the largest present supernova neutrino detectors. Super-Kamiokande
contains about 32,000 tons of  water in the inner volume. The SNO detector consists of 1,000 tons
of ultra-pure heavy water and about 1,700 tons of light water. Both the light and heavy water form active detector
volumes
for supernova neutrino detections, providing proton and deuteron targets for neutrino interactions.

Because of low, due to large distance, neutrino flux from candidate stars undergoing advanced 
stages of nuclear burning, very large detectors are needed in order to detect the signal.
Fortunately, there are proposals of two next generation water Cherenkov detectors for proton decay
searches, UNO \cite{uno} and Hyper-Kamiokande \cite{hk}, with 440,000 and 540,000 tons of water in fiducial volumes, respectively.
We consider here also two detectors employing liquid scintillator (LS) that will be capable of detecting supernova
neutrinos. These are KamLAND \cite{kamland}, already operating, and Borexino \cite{borexino}, to be operating soon.
The low energy threshold for  recoil electrons and
the scintillation light are their most important advantages over water Cherenkov detectors.

\subsubsection{Water Cherenkov detectors}

Water Cherenkov detectors have good efficiency of detection of recoil electrons with energy above 
a rather high threshold (about 5 MeV).
Recoil electrons are produced mainly in NC/CC interactions between neutrinos and electrons. However, almost all
interesting events will be produced in CC interactions between
electron anti-neutrinos and protons (inverse $\beta$ decay). The neutrino/anti-neutrino spectrum is closely related to
the electron/positron spectrum. Pair-annihilation neutrinos from considered last stages of pre-supernova star
cannot be detected  because of their low energy.
We want, however, to point out that despite this high energy threshold problem, we may detect the signal from
capture of neutrons produced in inverse $\beta$
decay, if some neutron absorbent material will be added to the pure water. The salt phase of the SNO experiment
proves that this is feasible. SNO detects $^{36}$Cl de-excition via an 8.6 MeV total energy gamma-ray cascade 
which is distinguishable from CC (single energetic particle Cherenkov)
events by angular isotropy measures.

The remaining flavors, $\nu_{\mu}$, $\nu_{\tau}$, $\bar{\nu}_{\mu}$ and $\bar{\nu}_{\tau} $, interact
only by neutral-currents, i.e. only elastically scatter off electrons.

\subsubsection{Liquid scintillator detectors}

The primary goal of the Kamioka Liquid scintillator Anti-Neutrino Detector (KamLAND) experiment is a search
for the oscillation of $\bar{\nu}_e$'s emitted from distant power reactors \cite{kamland}. The inverse $\beta$ decay reaction is
utilized to detect $\bar{\nu}_e$'s with energies above 1.8 MeV in 1,000 tons of liquid scintillator (LS). The detection of the $e^{+}$
and the 2.2 MeV gamma-ray from neutron capture on proton in delayed coincidence is a powerful tool for reducing
background.

The Borexino experiment main physics goal is the detection of the 0.86 MeV $\nu_e$'s from electron-capture
decay of $^7$Be in the Sun \cite{borexino}. The central volume of the detector contains 300 tons of liquid 
scintillator.
Neutrinos from the last stages of nuclear burning will interact in the LS via electron scattering and the
inverse $\beta$-decay.

\subsection{Reactions \& Cross-sections}

\subsubsection{ The $\bar{\nu}_e + p \rightarrow e^+ + n$ reaction}

Neutrino-proton scattering (often referred to as quasi-elastic neutrino-proton
scattering) is historically called inverse beta decay. The structure of nucleon
in neutrino-proton scattering at MeV energies is not important, because the 
momentum transfer between the two nucleons is small. We may approximate
the vector and axial-vector form factors with two constants: $F_V(0) = g_V = 1$
and $F_A(0) = g_A = 1.27$. The value of $g_A/g_V$ is determined from the measured
lifetime of the neutron.

Electron antineutrino quasi-elastic scattering on proton turns a proton into 
a neutron and produces positron
of $E_{e^{+}} = E_\nu - \Delta$ energy (at zeroth order in $1/M$), where 
$\Delta = M_n - M_p$ is the mass difference between the proton
and neutron. The  momentum of produced positron is 
$p_{e^{+}} = \sqrt{E^{2}_{e^{+}} - m_e^2}$.    

The standard expression for the total cross section is,
\begin{eqnarray}
\sigma_{tot}  =
\sigma_0\; (g_V^2 + 3 g_A^2)\; E_{e^{+}} p_{e^{+}}
 =  0.0952 \left(\frac{E_{e^{+}} p_{e^{+}}}{1 {\rm\ MeV}^2}\right)
\times 10^{-42} {\rm\ cm}^2\, .
\label{eq:sigtot0}
\end{eqnarray}

The constant $\sigma_0$ includes the energy-independent
inner radiative corrections \cite{beacombeta} :

\begin{eqnarray}
\sigma_0 = \frac{G^2_F cos^2\theta_C}{\pi}(1 + \Delta^R_{inner}),
\label{eq:sigma0}
\end{eqnarray}

where $\Delta^R_{inner} \simeq 0.024$, and the Cabbibo angle $cos\theta_C=0.974$. 

The inverse neutron $\beta$-decay, $\bar{\nu}_e + p \rightarrow e^+ + n$, is 
the reaction giving the largest yield for the detection
of pre-supernova neutrinos. The large cross section, low energy threshold,
delayed coincidence between positron's annihilation and neutron capture signals,
abudance of target protons makes this reaction the most promising.

We calculate the signal, which is the number of produced neutrons from this 
reaction in all the considered experiments.

\subsubsection{Neutrino-Electron Scattering}

\begin{table}
\begin{tabular}{ c c c c c }
\hline
Reaction & $c_L$ & $c_R$ & $c^{2}_L + \frac{1}{3} c^{2}_R$ & $ \frac{1}{2} c_L c_R $ \\
\hline
$\nu_e e^{-}$ & $1/2 + sin^2\theta_W$ & $sin^2\theta_W$ & 0.5525 & 0.0845 \\
$\bar{\nu}_e e^{-}$ & $sin^2\theta_W$ & $1/2 + sin^2\theta_W$ & 0.2317 & 0.0845 \\
$\nu_{\mu,\tau} e^{-}$ & $-1/2 + sin^2\theta_W$ & $sin^2\theta_W$ & 0.0901 & -0.0311 \\
$\bar{\nu}_{\mu,\tau} e^{-}$ & $sin^2\theta_W$ & $-1/2 + sin^2\theta_W$ & 0.0775 & -0.0311 \\
\hline
\end{tabular}
\caption{Coefficients that appear in neutrino-electron scattering cross section \cite{fukugita}.}
\label{coefficients}
\end{table}

Neutrino-electron scattering produces recoil electrons with kinetic energy from
zero up to the kinematic maximum. The laboratory differential cross section for
the $\nu_xe^{-}$ scattering is of the form:

\begin{eqnarray}
\frac{d\sigma}{dT'_e} = \frac{2 G^{2}_F m_e}{\pi} \left[ c^{2}_L + 
c^{2}_R \left( \frac{E_{\nu}-T'_e}{E_{\nu}} \right)^{2} - c_{L} c_{R} 
\frac{m_e}{E_{\nu}}\frac{T'_e}{E_{\nu}} \right]
\label{eq:nuediff}
\end{eqnarray}

where $T'_e = E'_e - m_e$ is the recoil electron kinetic energy.

In our rate calculation we approximate by integrating eq. (\ref{eq:nuediff}) over all electron recoil 
energies. The total cross section is 

\begin{eqnarray}
\sigma = \int_{0}^{T'_{max}} \frac{d\sigma}{dT'_e} \, dT'_e =  \frac{2 G^{2}_F m_e E_{\nu}}{\pi} \left[ c^{2}_L + 
\frac{1}{3} c^{2}_R  - \frac{1}{2}c_{L} c_{R} \frac{m_e}{E_{\nu}} \right]
\label{eq:nuetot}
\end{eqnarray}

where $c_L, c_R$ coefficients depend on the neutrino species considered 
(Table \ref{coefficients}). Total scattering cross section is calculated without radiative 
corrections.

The elastically-scattered electrons will have kinetic energies of a few MeV. These relativistic
electrons will be difficult to detect in any Cherenkov detector, because of high energy threshold. 
However, the next-generation liquid-scintillator
Borexino detector will have low target threshold of approximately 0.25 MeV, mainly due to progress
in radiopurity research \cite{zuzel}. It is possible that KamLAND experiment will improve radiopurity to detect
low energy recoil electrons in future, so we
calculate signal from elastic processes in both detectors.

\subsubsection{Interactions with heavy water}

The Sudbury Neutrino Observatory (SNO) employs inelastic neutrino-deuteron
scattering to study the solar neutrino flux. The total $^{8}B$ solar neutrino
flux is determined from observation of the following reactions :

\begin{eqnarray*}
\nu_{e}+d\rightarrow p+p+e^{-} & (CC),\\
\nu_{x}+d\rightarrow p+n+\nu_{x} & (NC),\\
\nu_{x}+e^{-}\rightarrow \nu_{x}+e^{-} & (ES),
\end{eqnarray*}

where $x=e,\mu$ or $\tau$. The pre-supernova anti-neutrinos of all flavors 
can interact with deuterons and electrons through the following additional 
reactions :

\begin{eqnarray*}
\bar{\nu}_{e}+d\rightarrow n+n+e^{+} & (inverse\,\beta), \\
\bar{\nu}_{x}+d\rightarrow p+n+\bar{\nu}_{x} & (NC), \\
\bar{\nu}_{x}+e^{-}\rightarrow \bar{\nu}_{x}+e^{-} & (ES).
\end{eqnarray*}

By observing the charged-current interactions (CC and $inverse\,\beta-decay$) 
only electron neutrino (anti-neutrino) flux is measured.
The NC reaction of any flavour neutrino on deuteron breaks up the deteron 
and produces a proton and a neutron. The NC cross-section is the same for all
neutrino flavours, thus one can determine the total flux of all active neutrino 
and anti-neutrino flavors above an energy threshold of 2.2 MeV. 
The cross sections for $\nu$-d reactions are of primary importance in the
analysis of SNO data, it is motivation for theoretical calculations employing
nuclear physics and effective field thory. 
We use in our paper the cross section values for the deuteron break-up
interactions from the tables provided by Kubodera and Nozawa \cite{kubodera}.
In all cases of anti-neutrino reactions with light or heavy water only the neutron is detected in the final
state. The ES reaction produces low energy recoil electron mainly below energy
threshold, so we do not consider further this reaction.

\subsubsection{Averaged cross section}

The spectrum-averaged cross section is

\begin{eqnarray}
\bar{\sigma}_\alpha = \int_0^{\infty} \sigma(E) \lambda_\alpha(E) dE
\end{eqnarray}

where $\lambda_\alpha(E)$ is the $normalized$ spectrum for neutrinos produced by
the pair-annihilation in pre-supernova star at C, Ne, O and Si burning
phase ($\alpha$ = C, Ne, O \&  Si). The integration over the spectrum of
incoming neutrino energies, E, was performed numerically and the results are shown in
Table~\ref{table_cross}. We use the total cross sections $\sigma(E)$ for the most important
interactions in considered detectors as described in previous sections.

\begin{table*}

\begin{tabular}{ c c c c c c }
\hline
$Reaction$ & $E_{th}\,[MeV] $ & $ \bar{\sigma}_{Si}\,[cm^{2}]$ & $\bar{\sigma}_O\,[cm^{2}]$ & $\bar{\sigma}_{Ne}\,[cm^{2}]$ & $\bar{\sigma}_C\,[cm^{2}]$  \\
\hline 
$\bar{\nu}_e + p \rightarrow e^+ + n$ & $1.8$ & $6.80 \cdot 10^{-44} $ & $3.74 \cdot 10^{-45} $ & $9.07 \cdot 10^{-46} $ & $4.88 \cdot 10^{-49} $ \\
$\bar{\nu}_e + d \rightarrow e^+ + n + n$ & $4.0$ & $1.22 \cdot 10^{-46} $ & $4.38 \cdot 10^{-50} $  & $4.64 \cdot 10^{-52} $ & ---\\
$\nu_x + d \rightarrow \nu_x + p + n$ & $2.2$ & $1.68 \cdot 10^{-45} $  & $1.64 \cdot 10^{-47} $  & $1.63 \cdot 10^{-48} $  & $ 4.76 \cdot 10^{-54} $\\
$\bar{\nu}_x + d \rightarrow \bar{\nu}_x + p + n$ & $2.2$ & $1.41 \cdot 10^{-45} $ & $1.20 \cdot 10^{-47} $  & $1.19 \cdot 10^{-48} $ & $3.27 \cdot 10^{-54} $\\
$\nu_e + e^{-} \rightarrow \nu_e + e^{-}$ & $0.0$ & $1.76 \cdot 10^{-44} $ & $1.04 \cdot 10^{-44} $  & $9.07 \cdot 10^{-45} $ & $6.40 \cdot 10^{-45} $\\
$\bar{\nu}_e + e^{-} \rightarrow \bar{\nu}_e + e^{-}$ & $0.0$ & $6.79 \cdot 10^{-45} $ & $4.05 \cdot 10^{-45} $  & $3.49 \cdot 10^{-45} $ & $2.45 \cdot 10^{-45} $\\
$\nu_{\mu,\tau} + e^{-} \rightarrow \nu_{\mu,\tau} + e^{-}$ & $0.0$ & $3.07 \cdot 10^{-45} $ & $1.95 \cdot 10^{-45} $  & $1.72 \cdot 10^{-45} $ & $1.26\cdot 10^{-45} $\\
$\bar{\nu}_{\mu,\tau} + e^{-} \rightarrow \bar{\nu}_{\mu,\tau} + e^{-}$ & $0.0$ & $2.63 \cdot 10^{-45} $ & $1.68 \cdot 10^{-45} $  & $1.49 \cdot 10^{-45} $ & $1.10 \cdot 10^{-45} $\\
\hline

\end{tabular}
\caption{Spectrum-averaged cross sections for interactions of pre-supernova neutrinos in the light- and
heavy-water Cherenkov detectors
 and liquid scintillator detectors. $E_{th}$ is the neutrino energy threshold for a given reaction;
 $\nu_x$ stands for $\nu_e$, $\nu_\mu$ and $\nu_\tau$; $\bar{\nu}_x$ stands for $\bar{\nu}_e$, $\bar{\nu}_\mu$ and $\bar{\nu}_\tau$.}
\label{table_cross}

\end{table*}

\section{The event rate and the background signal in selected detectors}

The number of incoming particles interacting (per day) with the target  is equal to the
product of the spectrum averaged
cross section $\bar{\sigma}_\alpha$, the number $N$ of target atoms or electrons,  the total intensity of the flux
per day $\phi_\alpha$,
and the fraction $f$ of total flux that is interacting with a given target.
The reaction rate $r$ per day is written as:

\begin{eqnarray}
r \, [day^{-1}] = f \cdot \; \bar{\sigma}_\alpha \, [cm^2] \cdot N \cdot \phi_\alpha \, [ cm^{-2} \, day^{-1}].
\label{r}
\end{eqnarray}

For anti-neutrinos from silicon burning stage in 1kt water Cherenkov detector we have :
$ \bar{\sigma}_{Si} \, = \, 6.8 \cdot 10^{-44} \, [cm^2], N \, = \, 6.69 \cdot 10^{31}, \,
\phi_{Si} \, = \, 7.6 \cdot 10^{11} \, [ cm^{-2} \, day^{-1}] $ and, from Table~\ref{f_factor},  $ f \, = \, 0.363$.

\begin{figure}
\includegraphics[angle=0,scale=.5]{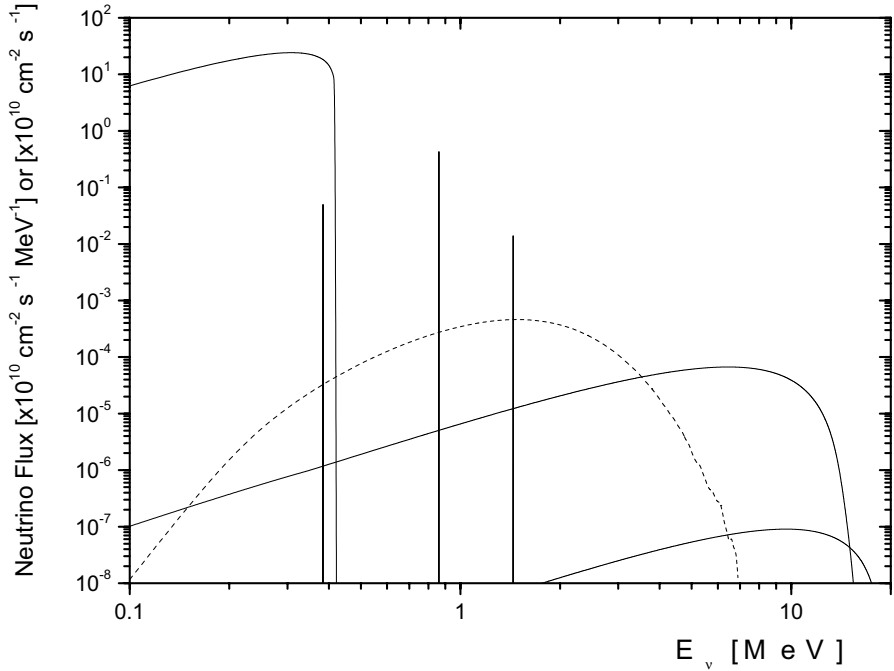}
\caption{The standard solar neutrino spectrum (BP2000, \cite{bahcall}) for pp fusion reactions in the Sun (solid lines)
and the spectrum of pair-annihilation neutrinos emitted by a 20~M$_\odot$ star during silicon burning stage (dashed line).
Star is located at a distance of 1~kpc.}
\label{solar}
\end{figure}

The number of target particles is easy to calculate if the mass and
the chemical composition of material in the fiducial volume are known. We obtain the total flux per day from
the pre-supernova star 1~kpc away by dividing the luminosity $L_\nu$ (Table~\ref{table2}) by the average
neutrino energy (Table~\ref{f_factor}) and the surface area of a R=1~kpc radius  sphere.
The spectrum of the total neutrino flux from the silicon burning stage is compared with that for solar neutrinos
in Fig.~\ref{solar}. The fraction of total flux that interact, $f$, depends on the burning stage and the type of
interactions in the detector.
All these expected properties are shown in Table~\ref{tbl:events}. Here the most important is the final result
of our calculation which is the event rate per day.

The equation (\ref{r}) gives for the silicon burning neutrinos 1.27 neutrons/day/kiloton
of water for a star 1~kpc away. Event rates for large detectors are summarized in Table~\ref{tbl:events}.
Super-Kamiokande is the best currently working detector for a such an observation, with 41 events per day, but
needs a modification for making neutron detection possible. The same modification would be required in the
light water volume of the SNO detector, allowing it to detect 2.2 events/day.

\begin{table}
\begin{tabular}{ r c c c c l } \hline
Detector & Mass     &  Reactions    &  Number of  & Flux at 1~kpc 		& Event rate		\\
         & $[kton]$ &               &  Targets    & $[cm^{-2}\,day^{-1}]$ 	& $[day^{-1}]$    	\\
\hline
Borexino & 0.3 $(C_9H_{12})$ &$\bar{\nu}_e + p \rightarrow e^+ + n$    				  & $1.80 \cdot 10^{31}$ & $2.8 \cdot 10^{11}$ & 0.34\\
         &  &$\nu_e + e^{-} \rightarrow \nu_e + e^{-}$                                & $9.92 \cdot 10^{31}$ & $2.8 \cdot 10^{11}$ & 0.49\\
         & &$\bar{\nu}_e + e^{-} \rightarrow \bar{\nu}_e + e^{-}$                  	  & $9.92 \cdot 10^{31}$ & $2.8 \cdot 10^{11}$ & 0.19 \\
         & &$\nu_{\mu,\tau} + e^{-} \rightarrow \nu_{\mu,\tau} + e^{-}$                   & $9.92 \cdot 10^{31}$ & $1.0 \cdot 10^{11}$ & 0.03 \\
         & &$\bar{\nu}_{\mu,\tau} + e^{-} \rightarrow \bar{\nu}_{\mu,\tau} + e^{-}$       & $9.92 \cdot 10^{31}$ & $1.0 \cdot 10^{11}$ & 0.026 \\
\hline
KamLAND  & 0.2 $(C_9H_{12})$ &$\bar{\nu}_e + p \rightarrow e^+ + n$    			  & $8.55 \cdot 10^{31}$ & $2.8 \cdot 10^{11}$ & 1.6\\
         & 0.8 $(C_{12}H_{26})$ &$\nu_e + e^{-} \rightarrow \nu_e + e^{-}$                & $3.43 \cdot 10^{32}$ & $2.8 \cdot 10^{11}$ & 1.7\\
         & &$\bar{\nu}_e + e^{-} \rightarrow \bar{\nu}_e + e^{-}$                  	  & $3.43 \cdot 10^{32}$ & $2.8 \cdot 10^{11}$ & 0.65\\
         & &$\nu_{\mu,\tau} + e^{-} \rightarrow \nu_{\mu,\tau} + e^{-}$            	  & $3.43 \cdot 10^{32}$ & $1.0 \cdot 10^{11}$ & 0.11\\
         & &$\bar{\nu}_{\mu,\tau} + e^{-} \rightarrow \bar{\nu}_{\mu,\tau} + e^{-}$       & $3.43 \cdot 10^{32}$ & $1.0 \cdot 10^{11}$ & 0.09\\
\hline
SNO      & 1.7 $(H_2O)$&$\bar{\nu}_e + p \rightarrow e^+ + n$    		   &$1.14 \cdot 10^{32}$  & $2.8 \cdot 10^{11}$  & 2.2   \\
         & 1 $(D_2O)$  &$\bar{\nu}_e + d \rightarrow e^+ + n + n$                  &$6.00 \cdot 10^{31}$  & $2.8 \cdot 10^{11}$  & 0.004\\
         & &$\nu_x + d \rightarrow \nu_x + p + n$                             	   &$6.00 \cdot 10^{31}$  & $3.8 \cdot 10^{11}$  & 0.038\\
         & &$\bar{\nu}_x + d \rightarrow \bar{\nu}_x + p + n$                      &$6.00 \cdot 10^{31}$  & $3.8 \cdot 10^{11}$  & 0.032\\
\hline
Super-K  & 32 $(H_2O)$&$\bar{\nu}_e + p \rightarrow e^+ + n$    				  &  $2.14 \cdot 10^{33}$  & $2.8 \cdot 10^{11}$ & 41 \\
UNO	 & 440 $(H_2O)$&$\bar{\nu}_e + p \rightarrow e^+ + n$    				  &  $2.94 \cdot 10^{34}$  & $2.8 \cdot 10^{11}$ & 560\\
Hyper-K  & 540 $(H_2O)$&$\bar{\nu}_e + p \rightarrow e^+ + n$    				  &  $3.61 \cdot 10^{34}$  & $2.8 \cdot 10^{11}$ & 687\\
\hline

\end{tabular}
\caption{Event rate per day in selected neutrino detectors from silicon burning stage in
neutrino-cooled star at distance of 1~kpc.
\label{tbl:events}}
\end{table}

To make detection of pre-supernova anti-neutrinos feasible we propose to supplement
the existing and future water Cherenkov detectors with some
addition of NaCl or GdCl$_3$
(see the proposition made by J.F. Beacom and M.R. Vagins {\it GADZOOKS!} \cite{gadzooks})
to the water.
Neutrons produced in inverse
$\beta$ decay are captured by Gd or Cl nuclei producing  high energy (above 8 MeV) gamma cascades.
The addition of neutron absorbers is very important, because in a~pure light water neutrons are captured by
protons and produce not detectable 2.2~MeV gamma cascades, e.g. energy threshold in SK is about 5~MeV.
This modification has been proved to work in the salt phase of solar neutrino detection
in the SNO detector. The addition of NaCl to the kiloton of heavy water increased the neutron
capture efficiency and the associated Cherenkov light~\cite{salt}. Neutron capture efficiency is
different in every experiment and has to be calibrated with neutron source, as e.g. $^{252}$Cf in SNO.

Background neutron events are disturbing pre-supernova neutrino detection, and 
thus should be carefully studied in every detector before the addition of the 
neutron absorber. The average rate of background neutron production from 
activity in the D$_2$O region is 0.72 neutrons per day in the SNO detector 
\cite{salt}. These background events are produced mainly by deuteron 
photodisintegration (not relevant for a light water detectors) and due to the
external-source neutrons. 
The background of neutrons in spallation products from 
energetic muons can be suppressed by a~veto following the muon \cite{kamland}.
Long-lived muon induced isotopes, which decay by the production of a neutron 
and a beta (e.g. $^8He$, $^9Li$ and $^{11}Li$), can simulate the $\bar{\nu}_e$ 
coincidence signature. Thus, the study of muon induced spallation products 
in the neutron absorber nuclei is important. 
Other major sources of the neutron background are : ($\alpha$,$n$) interactions,
fission of radioactive impurities, interactions of 
atmospheric and fission anti-neutrinos (both natural and from reactors)
with protons. The neutrino flux from nuclear power plants depends strongly 
on location of nearby ones, and this factor has to be included in projects 
of future pre-supernova detectors.  

\section{Importance of pre-supernova neutrinos detection}

Our results for $20~M_\odot$ pre-supernova star show that neutrinos
(particularly $\bar{\nu}_{e}$)
from last stages of nuclear burning in such a star are possible to
detect.
Future detectors (UNO, Hyper-Kamiokande) will be able to cover significant fraction of the Galaxy.
However, when we take into account relatively low, from typical experiment
lifetime point of view,
local {\it core-collapse} SN rate, of~1~per $\sim$100~yrs or less
 (Table~I of~\cite{odrzywolek}, \cite{beacom_mezzacappa} for overview, see however \cite{obscuredsn}),
only projects of long standing, namely nucleon decay and neutrino observatories, have chance
to be operating during such an event.
At present it is difficult to reliably estimate probability of
successful neutrino-cooled start detection because of 
(1) uncertainties of local supernova rate,
(2) unexplored yet variety of massive stars and their models,
(3) unknown details of detection technique, especially neutron
background level.
Making optimistic assumption of 10 events/day detectable and clearly 
distinguishable from background,
we get (using Galaxy model \cite{galaxy_model} and Si burning
event rate for 20 M$_\odot$ star from Table~\ref{tbl:events}) 1.7\% Galaxy coverage
(i.e. up to 2~kpc from Earth) for Super Kamiokande. Next generation
detectors UNO and Hyper Kamiokande cover significant fraction
30\% (7.5~kpc) and 37\% (8.2~kpc) of Galaxy disk stars, respectively.
The highest Galaxy supernova rate estimate of 1 per 10 years
\cite{piran}
gives (under above assumptions) successful detection probability
of 3\% (SK), 60\% (UNO) and 75\% (HyperK) during 20 years
of uninterrupted observation.

Benefits given by the detection of the pre-supernova neutrino signal can be divided
into two sets: (1) those related to possible prediction of subsequent supernova event and (2)
insight into processes related to deep interior of massive stars prior to its death.

\subsection{Supernova prediction?}

Supernova event is an unpredictable phenomenon. Astronomers await nearby supernova for 400 years.
Therefore, many of them  speculate on the likely next Galaxy event. The list of candidates includes
Betelgeuse, Mira Ceti, Antares, Ras Algheti, $\gamma^2$ Vel, Sher25 and Eta Carinae. Unfortunately,
the surface of these stars is unaffected by (possibly) dramatic nuclear and hydrodynamical processes deep inside.
This is effect of $\tau\!\sim\!10000$ years timescale of hydrogen envelope \cite{heger}.
No electromagnetic observation at any wavelength and resolution can help. Only neutrinos carry
informations on the current state of the stellar core. That's why only neutrinos could warn us before
the supernova event. The case of SN1987A~\cite{woosley} unexpectedly revealed this fact.
The silicon-burning neutrinos, although carry only about 1\% energy compared to the main supernova neutrino burst,
are possible to detect, as we showed in previous sections (cf. Table~\ref{tbl:events}). The information about
an incoming supernova
is transmitted around 2 days\footnote{
For a 20~M$_\odot$ star. This time strongly depends on the stellar mass (cf.~\cite{heger}, Table~I),
and is in the range 0.7 - 18 days.
}
earlier. The early warning would provide an additional time which may be crucial for preparation of all
available observational techniques,
including gravitational radiation detectors, and would allow us to be ready for the main
neutrino burst from collapse and protoneutron star cooling.

In a very favorable case of a close star, much less than 1~kpc away\footnote{
A 20~M$_\odot$ pre-SN star at Betelgeuse distance (Betelgeuse is actually a 15 M$_\odot$ red giant at
a distance of 185~pc)
which just entered the oxygen burning stage would produce in
HyperK 45 neutrons/day during 6 months before
its explosion.
}
with operating megaton-scale neutrino
observatory modified by addition of appropriate neutron absorber,
we could expect detection of oxygen- and neon-burning neutrinos a few months before the explosion.
The detection, however, would be more difficult than silicon-burning neutrinos, mainly due to lower 
luminosities L$_\nu$ (Table~\ref{table2}), lower average energies $\mathrm{E}_\nu$ (Fig.~\ref{spectra}) and
smaller cross-sections (Table~\ref{table_cross}).

Let's note, that the above speculations are no longer valid if the neutrino detection is {\it offline}.
Realtime  ({\it on-line}) data analysis is strongly preferred from this point of view.

\subsection{Astrophysical importance of Si burning neutrinos}

The aim of our work is to show the feasibility of pair-annihilation neutrinos detection. We did not discuss 
the calculations of the neutrino luminosities, but actually the silicon burning is very complicated and
''potentially numerically unstable stage'' (\cite{heger}, sec. IV A-4) of stellar modeling, mainly due to
similar timescales of nuclear burning and convection. The behaviour of spherically symmetric
models employing the mixing-length convection appears to be completely different
from 2D hydrodynamic models (\cite{arnett}, Epilogue).
Thus any observational data, even few
detected events may be very important to constrain theoretical models. In a favorable
situation of a close star with new-generation observatories we should be able
to constrain the time-dependence of the neutrino flux and of the spectrum.

The Si-burning neutrino emission precedes a subsequent explosion event independently of
the actual stellar death scenario. The SN1987A confirmed standard supernova mechanism \cite{bethe}.
However, it is believed that at least some of the massive stars die in other ways. The most recent
research is concentrated on GRB-SN connection \cite{macfayden}, their relation to failed 
supernovae \cite{failedsn}
and the core rotation \cite{odrzywolek}.
It is not yet understood why some massive stars become supernovae, hypernovae or even GRBs.
The detection of pre-supernova Si-burning neutrinos together with the following observations of optical,
neutrino  and gravitational signals
from the supernova and the identification of the progenitor would establish the relation of
pre-supernova conditions and the explosion dynamics. Let's note, that in case of the supernova shrouded
in interstellar clouds, Si burning neutrinos carry exclusive information on the progenitor.

\subsection{Discussion}

This article  shows that detection of pre-supernova star neutrinos is a feasible new goal for neutrino astronomy.
Our simplified analysis may be extended to stars of different masses. The constant neutrino flux
can be replaced with more realistic time-dependent flux generated by
stellar evolution codes. The simple spectrum of pair-annihilation neutrinos can be augmented with detailed
plasma- and weak-nuclear-neutrinos spectra. These refinements could change
the results somewhat, not affecting our predictions as to the detection feasibility considerably.
Clearly, the most important circumstance is the distance to the  next Galactic supernova. It is also, however,
the most indeterminable one.
Therefore, if we want to get results, we have to maximize our observation range.
Detectors like Hyper-Kamiokande, UNO, or even better should be operating at the ``time zero''.

\section{Acknowledgments}
We thank K. Grotowski and M. Wojcik for helpful discussions and advice. This work was partially supported by
the research grant from the Institute of Physics of Jagiellonian University, and through KBN grants 2 P03B 110 24
and PBZ-KBN-054/P03/02.

\end{document}